\documentclass[11pt,fleqn,twoside]{article}

\usepackage{amsmath}
\usepackage{amsthm}
\usepackage{amssymb}
\makeatletter
\newcommand{\prava}[1]{\small\it
\begin{flushleft}
Copyright \copyright \ 2000 by  #1
\end{flushleft}}

\newcommand{\name}[1]{\begin{flushleft}
                       \LARGE \bf #1
                       \end{flushleft}\vspace{-3mm}}

\newcommand{\Author}[1]{\begin{flushleft}
                       \it #1 \end{flushleft}}

\newcommand{\Adress}[1]{\begin{flushleft}
                       \it #1 \end{flushleft}}

\newcommand{\Date}[1]{\begin{flushleft}
                      \small  \it #1 \end{flushleft}}

\newcommand{\ehkol}{Author \ name}
\newcommand{\ohkol}{Article \ name}

\renewcommand{\@evenhead}{
\hspace*{-3pt}\raisebox{-15pt}[\headheight][0pt]{\vbox{\hbox to \textwidth 
{\thepage \hfil \ehkol}\vskip4pt \hrule}}}
\renewcommand{\@oddhead}{
\hspace*{-3pt}\raisebox{-15pt}[\headheight][0pt]{\vbox{\hbox to \textwidth 
{\ohkol \hfil \thepage}\vskip4pt\hrule}}}
\renewcommand{\@evenfoot}{}
\renewcommand{\@oddfoot}{}

\long\def\@makecaption#1#2{%
  \vskip\abovecaptionskip
  \sbox\@tempboxa{\small \textbf{#1.}\ \ #2}%
  \ifdim \wd\@tempboxa >\hsize
    {\small \textbf{#1.}\ \ #2}\par
  \else
    \global \@minipagefalse
    \hb@xt@\hsize{\hfil\box\@tempboxa\hfil}%
  \fi
  \vskip\belowcaptionskip}

\def\numberwithin#1#2{\@ifundefined{c@#1}{\@nocounterr{#1}}{%
  \@ifundefined{c@#2}{\@nocnterr{#2}}{%
  \@addtoreset{#1}{#2}%
  \toks@\@xp\@xp\@xp{\csname the#1\endcsname}%
  \@xp\xdef\csname the#1\endcsname
    {\@xp\@nx\csname the#2\endcsname
     .\the\toks@}}}}

\newcommand{\resetfootnoterule} {
  \renewcommand\footnoterule{%
  \kern-3\p@
  \hrule\@width.4\columnwidth
  \kern2.6\p@}
}

\makeatother

\numberwithin{equation}{section}

\makeatletter
\DeclareRobustCommand{\primfrac}[1]{%
  \PackageWarning{amsmath}{%
Foreign command \@backslashchar#1; %
\protect\frac\space or \protect\genfrac\space should be used instead%
  }
  \global\@xp\let\csname#1\@xp\endcsname\csname @@#1\endcsname
  \csname#1\endcsname
}
\makeatother

\begin{document}

\thispagestyle{empty}
\renewcommand{\ehkol}{A.M.\ Korostil}
\renewcommand{\ohkol}{Finite-Gap  
Solutions of the KdV Equation}

\begin{flushleft}
\footnotesize \sf
Journal of Nonlinear Mathematical Physics \qquad 2000, V.7, N~1,
\pageref{korostil_fp}--\pageref{korostil_lp}.
\hfill {\sc Article}
\end{flushleft}

\vspace{-5mm}

\renewcommand{\footnoterule}{}
{\renewcommand{\thefootnote}{}
 \footnotetext{\prava{A.M.\ Korostil}}}

\name{On the Calculation of Finite-Gap \\ 
Solutions of the KdV Equation}\label{korostil_fp}

\Author{A.M.\ KOROSTIL}

\Adress{Institute of Magnetism of NASU,
36 Vernadskii str., 252142 Kiev, Ukraine \\
e-mail: amk@imag.kiev.ua}

\Date{Received July 16, 1999; Revised September 11, 1999; Accepted
September 16, 1999}

\begin{abstract}
\noindent
A simple and general approach for calculating the elliptic finite-gap
solutions of the Korteweg-de Vries (KdV) equation is proposed.
Our approach is based on the use of the
finite-gap equations and the general representation of these solutions
in the form of rational functions of the elliptic Weierstrass function.
The calculation of initial elliptic finite-gap solutions is reduced to
the solution of the finite-band equations with respect to the parameters of the
representation. The time evolution of these solutions is described
via the dynamic equations of their poles, integrated with the help of the
finite-gap equations. The proposed approach is applied by calculating the
elliptic 1-, 2- and 3-gap solutions of the KdV equations.
\end{abstract}

\section{Introduction}
In accordance with the finite-band theory, the integrable Korteweg-de
Vries (KdV) equation
can be considered as the compatibility condition of the two auxiliary
linear differential matrix equations
$\partial_{x_{\pm}} {\boldsymbol \Phi}=
{\boldsymbol {\mathrm U}}_{\pm}$ ($x_+\equiv x,\,x_-\equiv t$,
$\partial_x\equiv d/dx$) with matrix operators
${\boldsymbol {\mathrm U}}_{\pm}$,
coefficients of which, as is well known (see \cite{im75,zmnp80,kr89}), are
expressed through solutions of the KdV equation. The finite-gap solutions
of the  KdV equation are solutions of the spectral problem for these
equations with a finite-gap spectrum of eigenvalues. These are expressed
through multi-dimensional Riemann theta functions with implicit parameters,
the evaluation of which is the special algebraic geometrical problem
(see \cite{kr89,db81,mu84,be94}).
The class of elliptic finite-gap solutions leads to the problem
of the reduction of $n$-dimensional theta-functions to the one-dimensional
Jacobi theta functions (see \cite{be94}).

In the framework of the spectral problem the finite-gap solutions
satisfy the matrix finite-gap equation of the form
$\partial_{\pm}{\boldsymbol \Psi}=
{\boldsymbol {\mathrm U}}_{\pm}{\boldsymbol \Psi}-{\boldsymbol \Psi}
{\boldsymbol {\mathrm U}}_{\pm}$, where ${\boldsymbol \Psi}$ is the matrix
components which are polynomial in the eigenvalue of the auxiliary
equations (such as in the case of the ``sine-Gordon'' equation \cite{fm83}).
Usually, solving  this equation is realized with help of the
well known Abel transformation with a subsequent solving of the inverse
Jacobi problem (see \cite{db81,mu84}).

However, in the case under consideration the finite-gap solutions
of the KdV equation are elliptic functions represented as rational
functions of the elliptic Weierstrass function ($\wp$-function)
\cite{ba55}. Solving the finite-gap equations in terms of $\wp$-function
can be reduced to solving simple algebraic equations. This yields
the straightforward manner for calculating the elliptic finite-gap
solutions in an explicit form.

We shown that in the initial time ($t=0$) the finite-gap equations
can be reduced to algebraic equations with respect to parameters of
the above mentioned rational functions. The computation of these
parameters gives all possible initial elliptic finite-gap
solutions in the form of linear combinations of $\wp$-functions with
shifted arguments.

In accordance with the KdV equation the time dependent elliptic
solutions are built as linear combinations of $\wp$-functions
with time dependent argument shifts ($\varphi_i$) (its poles)
under condition of its compatibility with corresponding
initial solutions. Their time evolution is determined by the poles
which satisfy the system of linked dynamic equations which follow from
the above mentioned auxiliary equations.

Using the finite-gap equations we shown that the dynamic system is
transformed in the system of independent differential equations of the
first order with separated variables of the form
$\partial_t \varphi_i=X_i(\varphi_i)$.
Here $X_i$-functions represent themselves roots of some polynomial
equations which follow from the finite-gap equations the order of which
equals the number of poles in the elliptic solutions.

This paper is organized as follows.
In Section~\ref{secA} the approach to the straightforward calculation of the
elliptic finite-gap solutions of the KdV equation,  based on the
auxiliary system of finite-gap  and dynamic equations, is formulated.
In Section~\ref{secB} this approach is applied to the calculation of the
initial elliptic 1-, 2- and 3-gap solutions of the KdV equation.
In Section~\ref{secC} the time evolution of these elliptic solutions is
investigated. We shown that the linked system of auxiliary dynamic
equations, for poles of the corresponding time dependent elliptic solutions,
can be integrated with the help of the finite-gap equations.

\section{Finite-band equations and general elliptic solutions} \label{secA}

In the class of elliptic functions (which we shall  denote as  $U(x,t)$)
with the $\wp$-functional representation under consideration, the finite-gap
solutions of the KdV equation can be considered as solutions of the finite-band
equations. The latter gives the compatibility condition of the finite-gap
and the general solutions of the auxiliary linear differential equations. These
finite-band equations represent a system of equalities,
obtained by equating coefficients of the power series in the
eigenvalue
$E$ of the
finite-gap and general solutions (see \cite{zmnp80}).

The finite band equations (which are also known as ``trace-formulae'')
can be written in the
form \cite{zmnp80,ko95}
\begin{gather}
A_{n+1}={\frac{(-1)^n}{2^{2n+1}}}\chi _{2n+1}(x,t) \,\,(n=0,1,\ldots),\,a_0=1
\label{trace}\\
A_n=\frac{1}{n!} \partial_z^n \left( \frac {\sqrt{
\sum_{n=0}^{2g+1}a_nz^{n}}}{\sum_{n=0}^gb_n(x,t)z^n}\right)\Bigg|_{(z=0)},
\,b_0=1
\label{A_n}
\end{gather}
($g$ is the number of a gaps in the spectrum of the eigenvalues $E$),
where
the
$\chi_n$-functions are determined by the recursion relation
\begin{equation}
\chi_{n+1}=\partial_x \chi_n+\sum_{k=1}^{n-1}\chi_k\chi_{n-k},\,
\chi_1=-U(x).\label{rec}
\end{equation}
Here $A_n$ and $\chi_n$ are coefficient functions of the power-series
expansion
in $E$ of the general and the finite-gap solutions of the auxiliary
equations, respectively. In view of (\ref{rec}) the $\chi$-functions have the
form of polynomials in the solutions $U(x,t)$ and its derivatives. As
is well
known, (\ref{trace}) is algebraicly solvable with respect to the coefficient
functions $b_n(x,t)$. Therefore, the finite-gap equations (\ref{trace})
are reduced to the system containing expressions for the coefficient functions
$b_n(x,t),\,n=(1,\ldots,g)$, which at $n \le g$ is a closed system of
equations for the elliptic finite-gap $U$-solutions.

The elliptic finite-gap solutions as double periodic functions of the
complex variable $z$ which admit the rational
functions
representation
namely the elliptic Weierstrass function $\wp$ \cite{ba55}. In view of
the equation (\ref{trace}), $U$ is determined by the formula
\begin{equation}
U(z,t)= \alpha \wp(z)+\sum_i \sum_{n_i=1}^{2} \Big\{ \frac{ \alpha_{n_i}(t)}
{(\wp(z)-h_i(t))^{n_i}} \Big\}+{\tilde R}(z,t)  \label{New0}
\end{equation}
with poles of the secondary order in $\wp \equiv \wp(z|\omega,\omega')$
($\omega$ and $\omega'$ are real and imaginary half-periods of the
$\wp$-function), where ${\tilde R}(z,t)$ means an odd function of $z$.

At the initial time ($t=0$) the expression (\ref{New0}) describes all
possible so-called initial elliptic finite-gap solutions, which in the
case of even functions ($\tilde R = 0$) under consideration, takes the form
\begin{equation}
U(z)= \alpha \wp(z) + \sum_i \sum_{n_i=1}^{2}
 \frac{\alpha_{n_i}}{(\wp(z) - h_i)^{n_i}},\label{New}
\end{equation}
that reduces the finite-gap equations (\ref{trace}) to simple algebraic
equations with respect to the $\alpha$- and $h_i$-parameters.

The time evolution of the elliptic finite-gap solutions (\ref{New0}) are
determined by the time dependence of their poles, which are described
with the help of the known dynamic auxiliary equation
\begin{equation}
\partial_t \Psi (x,t,E)=
\left (4\partial+3(U\partial_x+\partial_x U)\right ) \Psi (x,t,E),\label{da}
\end{equation}
$\Psi=\Psi(\{a_n)\},\{b_n\})$,
$b_n=b_n(U(z,t)$, ${U^{(n)}}(z,t))$ can be reduced to the system of
dynamic equations with respect to the poles of the $U$-functions.
This is a system of coupled differential equations of the first order
which, as will be shown below, can be reduced to  independent equations
with separated variables.

\section{A calculation of the initial elliptic solutions} \label{secB}

The calculation of the initial finite-gap elliptic solutions of the
KdV equation is based on the use of the finite-band equations
(\ref{trace}) in the representation of the rational functions (\ref{New}).
On equating the corresponding coefficients of the Laurent expansion
in $\wp$ of the left- and right-hand sides (\ref{trace}), yields  simple
algebraic equations which determine the parameters of the expression
(\ref{New}) for the initial elliptic finite-gap solutions of the KdV
equation.

\subsection{One-gap elliptic solutions} \label{secBA}

In the one-gap case the parameters of the elliptic solution $U(z)$ are
determined by the system of three finite-gap equations of the form
(\ref{trace}) at $n=\overline {0,2}$. A substitution in these equations,
of the explicit expressions (\ref{A_n}) for $A_n$ and polynomials in $U$
expressions, for the $\chi_n$-functions which follow from (\ref{rec}),
yields
the system
\begin{equation}
\begin{split}
&\frac{1}{2}a_1-b_1=-\frac{1}{2}U;  \\
&\frac{1}{2}\{ (a_2-{1\over 4}a_1^2)+2(b_1^2-b_2)-a_1b_1 \}=-\frac{1}{2^3}
\{U^2-U^{(2)}\};\\
&\frac{1}{3!} \{ ({3\over 8}a_1^3-{3\over 2}a_1a_2+{3\over 2}a_3)
+3({1\over 4}a_1^2-a_2)b_1+3a_1(b_1^2-b_2)\\
&\qquad +(12b_1b_2-4!b_3-6b_1^3) \}=\frac{1}{2^5}\{U^{(4)}-5U^{(1)^2}+6UU^{(2)}-
2U^3\}
\end{split}\label{AS1}
\end{equation}
in which $b_n|_{n \ge 2}=0$ (in view of the relation $b_n|_{n \ge g+1}=0$),
where $g$ is the number of gaps in the spectrum of the auxiliary linear
differential equation. Excluding $b_n$ from the system (\ref{AS1}) we can
obtain the equations
\begin{equation}
\begin{split}
b_2=0={}&{1\over 8}(3U^2-U^{(2)})+{1\over 4}a_1U+{1\over 2}a_2-
{1\over 8}a_1^2;\\
b_3=0={}&-{1\over 32}(U^{(4)}+10U^3-5U^{(1)^2}-10UU^{(2)})-{1\over
16}a_1(3U^2-U^{(2)}) \\
 &+ {1\over 16}U(a_1^2-4a_2)+{1\over 2}a_3+
{1\over 4}a_1a_2-{1\over 16}a_1^3,
\end{split}\label{b23}
\end{equation}
which is known \cite{zmnp80} as the one-gap ``trace formulae''. These
equations form a closed system which determines the initial elliptic
solutions (\ref{New}). Inserting the rational expression for $U(z)$
(\ref{New}) into the system (\ref{b23}) and equating coefficients of
the Laurent expansion in $\wp$ in the right-hand sides to zero, we obtain
algebraic relations which lead to the equalities
\begin{equation}
\begin{split}
1)\quad&\alpha=2\,\alpha_{1_i}=\alpha_{2_i}=0,\, a_1=0,\,
a_2=-{1\over 4}g_2E,\, a_3=-\frac{1}{4}g_3;\\
2)\quad&\alpha=2,\,\alpha_{(1,2)_i}=\frac{1}{4}\beta_{(1,2)},\,
a_1=0,\,a_2=\frac{1}{4}(11g_2-120\wp^2(\varphi_1)),\\
&a_3=-\frac{1}{4}g_3+4\wp(\varphi_1)(12\wp^2(\varphi_1)-g_2)-
4\wp'{}^2(\varphi_1).
\end{split}\label{az1}
\end{equation}
Here and below $\beta_1=12h^2-g_2,\,\beta_2=2h'{}^2$,
$\varphi_1$ is the argument of the function $h=\wp(\varphi_1)$
which satisfies the equation $(12h^2-g_2)^2=48 h h'{}^2,$
from which $\varphi_1= (2/3)\omega_i|_{i=(1,2,4)}$, where
$\omega_4=(\omega-\omega')$ (see also \cite{sm94}).

The two systems (\ref{az1}) determine  two types  of initial elliptic
solutions:
\begin{equation}
\begin{split}
1)\quad&U(z)=2\wp(z);\\
2)\quad&U(z)=2\wp(z)+2[\wp(z-\varphi_1)+\wp(z+\varphi_1)]-
4\wp(\varphi_1),
\end{split}\label{U1}
\end{equation}
where the first type is the known \cite{ba55} Lam\'e potential and the
second type is
a  new potential obtained in \cite{sm94}.

\subsection{2-gap elliptic initial solutions} \label{secBB}

Coefficients of the initial elliptic 2-gap solutions  are determined
by the system of the four finite-gap equations of the form (\ref{trace})
at $n=\overline {0,3}$. In analogy to the one-gap case, the explicit
form  can be obtained by substituting the expressions (\ref{A_n})
for $A_n$ and the expressions for $\chi_n$ (from (\ref{rec})) into
(\ref{trace}). In doing so, the first two equations, which coincide with
the first two equations of the system (\ref{b23}), are solvable with  respect
to $b_1$ and $b_2$. Excluding the latter from the fourth and fifth equation
and taking into account the equality $b_n|_{n \ge 3}=0$ we obtain
the finite-gap system
\begin{equation}
\begin{split}
b_3=0={}&\frac{1}{2^5}(16a_3 +8a_2U + 10U^3 -
5U'{}^2 - 2a_1U'' \\
& - 10UU'' + U^{(4)});\\
b_4=0={}&\frac{1}{2^7}(-16a_2^2 + 64*a_4 + 32a_3U + 24a_2U^2 + 35U^4 \\
& - 70UU'{}^2 - 8a_2U'' - 70U^2U'' + 21U''{}^2 \\
& + 28U'U^{(3)} + 14UU^{(4)} -  U^{(6)}).
\end{split}\label{b34}
\end{equation}
Inserting the rational expression (\ref{New}) for $U(z)$  into the
system (\ref{b34}) and equating coefficients of the Laurent expansion
in $\wp$, on the right- and left-hand sides, we obtain algebraic relations
which lead to the equalities
\begin{equation}
\begin{split}
1)\quad&\alpha=6,\,\alpha_{1_i}=\alpha_{2_i}=0,\,a_1=0,\,a_2=-\tfrac{21}{4}g_2,\\
\,&a_3=-\tfrac{27}{4}g_3,\,a_4=\tfrac{27}{4}g_2^2,\,a_4=-\tfrac{81}{4}g_2g_3;\\
2)\quad&\alpha=6,\,\alpha_{1_i}=\delta_{i,j}(12e_j^2-\tfrac{1}{2}g_2),\,
a_2=-7(9e_j^2+2\Lambda_j),\\
&a_3=18(3e_j^3-5e_j\Lambda_j),\,a_4=27(36e_j^4+16e_j^2\Lambda_j+3\Lambda_j^2),\\
&a_5=-54(36e_j^5-52e_j^3\Lambda_j-9e_j\Lambda_j^2),\\
&\Lambda_j=3e_j^2-\tfrac{1}{4}g_2;\\
3)\quad&\alpha=6,\,\alpha_{1_i}=(\delta_{i,j}+\delta_{i,k})(12e_i^2-g_2),\,a_2=161\Lambda_j-378e_j^2,\\
&a_3=531e_j\Lambda_j+108e_j^3,\,a_4=27(240\Lambda_j^2+1280e_j^2-1159e_j^2\Lambda_j),\\
&a_5=27(1594e_j\Lambda_j^2+8100e_j^5+120e_j^3\Lambda_j^2);\\
4)\quad&\alpha=6,\,\alpha_{1_i}=\tfrac{1}{4}\beta_1 \delta_{i,1},\,
\alpha_{2_i}=\tfrac{1}{4}\beta_2 \delta_{i,1},\\
&a_2=\tfrac{7}{2}(18 h^2 + \tfrac{7}{2}\alpha_1),\,a_3=\tfrac{9}{4}(24 h^3 - 25 h'{}^2) - \tfrac{34h}{2}\alpha_1),\\
&a_4=\tfrac{27}{4}(144 h^4 + 44 h h'{}^2 + 56 h^2\alpha_1+\tfrac{23}{4}\alpha_1^2),\\
&a_5=-\tfrac{27}{2}(144 h^5 + 6 h^2 h'{}^2- 74 h^3\alpha_1-\tfrac{41}{2} h'{}^2\alpha_1 -\tfrac{75h}{2}\alpha_1^2),
\end{split}\label{Ua2}
\end{equation}
where  $\varphi_2$ is determined by the equality
$h=\wp(\varphi_2)$ which satisfies the equation
$64h'{}^4+48hh'{}^2\alpha_1-\alpha_1^3=0.$
These four types of equalities (\ref{Ua2}) lead to the following four 
expressions
\begin{equation}
\begin{aligned}
1)\quad&U(z)=6\wp(z);\\
2)\quad&U(z)=6\wp(z)+2\wp(z+\omega_i)-2e_i;\\
3)\quad&U(z)=6\wp(z)+2\wp(z+\omega_i)+2\wp(z+\omega_k)
       -2(\wp(\omega_i)+\wp(\omega_j));\\
4)\quad&U(z)=6\wp(z)+2\wp(z-\varphi_2)+2\wp(z+\varphi_2))-
4\wp(\varphi_2).
\end{aligned}\label{U2}
\end{equation}
which describe all possible initial two-gap solutions of the KdV equations
(see \cite{ee95,sm94}). The first type coincides with the known
two-gap Lam\'e potential, while the second and third type
corresponds to the Treibich-Verdier potential \cite{ba55,tv91}.
The fourth solution is the new two-gap potential obtained
in \cite{sm94}.

\subsection{Initial elliptic three-gap solutions} \label{secBC}

The parameters of the initial elliptic three-gap solutions (\ref{New}) are
determined by the system of the finite-band equations (\ref{trace})
at $b_{n \le 3} \ne 0$ and $b_{n > 3}=0$. Using the expressions (\ref{A_n})
and (\ref{rec}) this system can be expressed through $U$-functions. In
doing so, the first three equations coinciding formally with
(\ref{AS1}) and are
solvable with respect to $\overline {b_1,b_3}$. Therefore,
taking into
account the equality $b_n|_{n \ge 4}=0$,  we can obtain the equation
\begin{align}
 b_4=0={}&-16a_2^2 + 64a_4 + 32a_3U + 24a_2U^2 +35*U^4 \notag\\
    &-70*UU'{}^2 - 8*a_2U''- 70*U^2U'' \notag\\
    &+ 21U''{}^2 + 28U'U'''+ 14UU^{(4)} - U^{(6)},\label{b4}
\end{align}
which is solvable with respect to parameters of the rational expression
$U(z)$ (\ref{New}). Substituting (\ref{New}) into the equation
(\ref{b4}) and equating coefficients of the Laurent expansion in $\wp$
with its right-hand side to zero we can obtain closed algebraic relations for
$\alpha$-parameters of the $U$-solution (\ref{New}).

It can be shown that there are four possible types of solutions for
the $\alpha$-parameters. The corresponding four types of initial elliptic
solutions can be written in the form
\begin{equation}
\begin{split}
1)\quad &U(z)=12\wp(z);\\
2)\quad &U(z)=12\wp(z)+2\wp(z+\omega_i)-2e_i,\,e_i=\wp(\omega_i);\\
3)\quad &U(z)=12\wp(z)+2(\wp(z+\omega_i)+\wp(z+\omega_j))-2(e_i+e_j);\\
4)\quad &U(z)=12\wp(z)+2(\wp(z+\varphi_3)+\wp(z-\varphi_3)) -4\wp(\varphi_3).
\end{split}\label{U3}
\end{equation}
Here the argument $\varphi_3$ is determined by the equation
\[
h^6+ {101\over 196}g_2h^4 + {29\over 49}g_3h^3- {43\over 784}g_2^2 h^2-
{23\over 196}g_2g_3h - \Bigl({1\over 3136}g_2^3 + {5\over 98}g_3^2\Bigr)=0
\]
where $h=\wp(\varphi_3)$.  The initial solutions 1 and 2, 3 are the
3-gap Lam\'e (\cite{ba55}) and generated on the 3-gap case
Treibich-Verdier \cite{tv91} potentials, respectively. The initial
elliptic solution 4 is the 3-gap generalization of the solution obtained
in \cite{sm94}.

Note that the above described algorithm is general and applicable
to computing  arbitrary initial elliptic $n$-gap  solutions of the
KdV equation.

\section{A dynamics of the elliptic finite-gap solutions} \label{secC}

The time dependent elliptic finite-gap  solutions of the KdV equation
have the general form of the rational functions of the $\wp$-function
(\ref{New0}), parameters of which are functions of time $t$.
These parameters are described by the system of the auxiliary dynamic
equation (\ref{da}) and the finite-gap equation (\ref{trace}).

By substituting the expression (\ref{New0}) for the elliptic
finite-gap solutions $U(z,t)$ into the KdV equation
$\partial_t U(z,t)=6U(z,t)U'(z,t)-U'''(z,t)$ and equating the
coefficients of the Laurent expansion in $\wp$, in its left- and right-hand
sides, lead to the known general formula
\begin{equation}
U(z,t)=2\sum_{i=1}^N \wp(z-\varphi_i(t))+C,\label{pot}
\end{equation}
in which the integer number $N$ and the constant $C$ are determined
by the condition  of the reduction of $U(z,t)$  to the corresponding initial
elliptic finite-gap solution at $t \to 0$. In the cases of the elliptic
1-, 2- and 3-gap solutions the numbers $N$  must provide
the reduction of the time dependent solutions (\ref{pot}) to
the corresponding initial elliptic solutions of the systems (\ref{U1}),
(\ref{U2}) and (\ref{U3}), respectively.

The substitution of the expression (\ref{pot}) into the KdV equation
reduces the problem of the time evolution of $U(z,t)$ to the time
evolution its poles. The latter is described by the system
\begin{equation}
\begin{split}
&\partial_t \varphi_i(t)=-12X_i(t)+C,\,\,
X_i(t)=\sum_{j=1,j \ne i}^{N-1}\wp(\varphi_i(t)-\varphi_j(t))
\,(g \ge 2),\\
&\partial_t \sum_{n=1}^N \wp(z-\varphi_i(t))=0\,\,(g=1),\\
\end{split}\label{Xd}
\end{equation}
which is a system of coupled equations. However, in view of the
symmetry properties of
the finite-gap equations, (\ref{trace}) determine $X_i(t)$ as function
$X_i(\varphi_i(t))$. Then the system (\ref{Xd}) can be transformed to
\begin {equation}
\int_{\varphi_{0i}}^{\varphi_i}\frac{d\varphi_i}
{X_i(h_i(\varphi_i))+C}=-12t,
\label{pd}
\end{equation}
which describe the time dynamics of the poles in the expression (\ref{pot}).
Here, initial values $\varphi_{0i} \equiv \varphi_i(0)$ are determined
from the expressions for initial elliptic finite-gap solutions.

The functions $X_i(\varphi_i)$ are determined by the finite-gap equations
(\ref{trace}) with $U$-functions in the form (\ref{pot}) as roots of
$N$th order polynomials in $X_i$. These polynomials are followed from
the algebraic equations which can be obtained by equating coefficients
of the Laurent expansion in $\wp$ of the left- and right-hand  sides of the
indicated equations (\ref{trace}).

Thus, the problem of time evolution of the elliptic finite-gap
solutions of the KdV equations is reduced to the solution of the
finite-band equations, with respect to the $X_i$-functions from
relations (\ref{pd}).
The proposed approach will now be applied in calculating 
the time evolution
of elliptic 1-, 2- and 3-gap solutions.

\subsection{Elliptic 1-gap solutions} \label{secCA}

The types of time dependent elliptic 1-gap  solutions of the KdV
equation are determined by the values of the number $N$ in the expression
(\ref{pot}). The condition of a coincidence in the general expression
(\ref{pot}) with the initial elliptic 1-gap solutions (\ref{U1})
at $t \to 0$, yields the values, namely  $N=1$ and $N=3$. The corresponding
elliptic 1-gap solutions have the form
\begin{equation}
\begin{split}
1)\quad&U(z,t)=2\wp(z-\varphi_1(t)),\\
2)\quad&U(z,t)=2\sum_1^3\wp(z-\varphi_i(t))	-4\wp(\varphi_1).
\end{split}\label{U1t}
\end{equation}
The time dependence of poles of the elliptic solutions 1 and 2 is determined
by the second equation of the system (\ref{Xd}) at $N=1$
and $N=2$, respectively. In accordance with the initial conditions
defined by the system (\ref{U1}), the solutions of the dynamic equation
have the form $\varphi_1=c_1t$ at  $N=1$ and
\[
\varphi_1(t)=c_1t,\,\varphi_2(t)=c_2t+\varphi_i^{(1)},\,
\varphi_3(t)=c_3t-\varphi_i^{(1)}
\]
at $N=3$. The substitution of these expressions into (\ref{U1t})
yields the following two expressions
\begin{equation*}
\begin{split}
1)\quad&U(z,t)=2\wp(z-c_1t);\\
2)\quad&U(z,t)=2\{\wp(z-c_1t)+\wp(z-c_2t+\varphi_i^{(1)})+
\wp(z-c_3t-\varphi_i^{(1)})-4\wp(\varphi_1),
\end{split}
\end{equation*}
which describes  two possible types of the elliptic  1-gap solutions
in the form of superpositions of one and three independent traveling
waves, respectively.

\subsection{Elliptic 2-gap solutions} \label{secCB}

The possible types of the time dependent elliptic 2-gap solutions of
the KdV equation (\ref{pot}) are determined by the values $N$ which can be
obtained from the compatibility condition between (\ref{pot}) and (\ref{U2})
as $t \to 0$. Under this condition the number $N$ equal 3, 4 and
5 in the formula (\ref{pot}).

1. The time dependent 2-gap elliptic solution  corresponding to the
initial condition 1 in the system (\ref{U2}), which is described by the
expression (\ref{pot}) at $N=3$, has the form
\begin{equation}
U(z,t)=2\sum_{i=1}^3\wp(z-\varphi_i(t)).\label{U21t}
\end{equation}
The time evolution of poles $\varphi_i(t)$ is described
by the equation (\ref{pd}) in which
$X_i=\sum_{j \ne i,j=1}^2 \wp(\varphi_i-\varphi_j)$.
Under initial conditions, the lower limits of the integration
in (\ref{pd}) are $\varphi_{0i}=0$.

To comput the $X_i$-function we substitute the
expression (\ref{U21t}) into the finite-gap equation (\ref{b23}).
Then, equating coefficients of the Laurent expansion in $\wp$ of the 
left- and right-hand sides, we obtain algebraic equations which are
reduced to the polynomial equation
\begin{equation}
X^3+c_2X^2+c_1X+c_0=0,\label{X3}
\end{equation}
where $c_n=c_n(\varphi),\,n=\overline {1,3}$ (here and below the
subscript $i$ of $\varphi$ is omitted).
Three solutions of (\ref{X3}) describe three unknown functions
$X_i(\varphi_i),\,i=\overline {1,3}$.
The dependence of the coefficient functions $c_n$ on $\varphi$  is
described by the expressions
\begin{equation*}
\begin{split}
&c_0=-\frac{36}{125}g_3 - \frac{2}{125}a_2 - \frac{1}{250}(149g_2+76a_2)h
+\frac{1}{400}(g_2+4a_2)\Bigl(\frac{\beta_1}{h'}\Bigr)^2;      \\
&c_1=\frac{29}{250}g_2 + \frac{8}{125}a_2;       \\
&c_2=-\frac{42}{125}h + \frac{3}{800}\Bigl(\frac{\beta_1}{ h'}\Bigr)^2, \\
\end{split}
\end{equation*}
where  $\beta_1=12h^2-g_2$ and $h=\wp(\varphi)$ so that $X_i$ depend
on $\varphi$ through the $h$-function.

2. The time dependent elliptic 2-gap solution corresponding to the
condition  2 in the system (\ref{U2}) is described by the expression
(\ref{pot}) at $N=4$ and has the form
\begin{equation}
U(z,t)=2\sum_{i=1}^4\wp(z-\varphi_i(t))-2\wp(\varphi_{04}).\label{U22t}
\end{equation}
The time evolution of poles $\varphi_i(t)$ are described via the function
$X_i=\sum_{j \ne i,j=1}^3 \wp(\varphi_i-\varphi_j)$.
In view of the initial conditions,  the lower limits of an integration
in (\ref{pd}) are $\varphi_{0i}|_{i \le 3}=0$ and $\varphi_{04}=\omega_j$.

For computing the $X_i$-functions we substitute the
expression (\ref{U22t}) into the system (\ref{b34}). Then, equating
coefficients of the Laurent expansion  of the  left- and
right-hand sides, we obtain the equation
\begin{equation}
X^4+c_3X^3+ c_2X^3+c_1X+c_0=0,\label{X4}
\end{equation}
solutions of which describe  the functions
$X_n(\varphi),\,n=\overline {1,4}$. The dependence of the
coefficient functions $c_n$ on $\varphi$  are  described by the
expressions
\begin{equation}
\begin{split}
c_0={}&-{\tilde m}^0_1g_2^2 +{\tilde m}^0_2g_2a_2 +{\tilde m}^0_3a_2^2-{\tilde m}^0_4a_4;\\
c_1={}&{\tilde m}^1_1g_3 +{\tilde m}^1_2a_3 +{\tilde m}^1_3g_2h +{\tilde m}^1_4a_2h 
    - {\tilde m}^1_5g_2\Big(\frac{\beta_1}{h'}\Big)^2 -
 {\tilde m}^1_6a_2\Big(\frac{\beta_1}{h'}\Big)^2;\\
c_2={}&-{\tilde m}^2_1g_2 -{\tilde m}^2_2a_2;\,c_3=-{\tilde m}^3_1h + {\tilde m}^3_2\Big( \frac{\beta_1}{h'}\Big)^2.\\
\end{split}.\label{CW4}
\end{equation}
Here ${\tilde m}^i_j$ denotes numerical parameters which have the  form
of some rational fractions.

3. The time dependent elliptic 2-gap  solutions, corresponding to
the initial conditions 3 and 4 in the system (\ref{U2}) which are described
by the expression (\ref{pot}) at $N=5$, have the form
\begin{equation}
U(z)=
2\sum_{i=1}^5\wp(z-\varphi_i(t))-2(\wp(\varphi_{04})+\wp(\varphi_{05})).
\label{U23t}
\end{equation}
In view of the initial conditions, the lower limits of the integration in
(\ref{pd}) have the common values
$\varphi_{0i}|_{i \le 3}=0$ and
$\varphi_{04}=\omega_i,\,\varphi_{05}=\omega_j$ at the condition 3 and
$\varphi_{04}=-\varphi_{05}=\varphi_2$ at the condition 4.

For computing the $X_i$-functions we substitute the
expression (\ref{U23t}) into the system (\ref{b34}). Then, equating
coefficients of the Laurent expansion in $\wp$ of the left- and right-hand
sides, we obtain the equation
\[
X^5+c_3X^3+c_2X^2+c_1X+c_0=0,\label{X5}
\]
solutions of which describe functions
$X_n(\varphi),\,n=\overline {1,5}$. Coefficient functions $c_n$ are
described by the expressions
\begin{equation*}
\begin{split}
c_0={}&\sum_{i=1}^5(m^0_ig_3+n^0_ia_3)F_i(\varphi) 
 + m^0_6g_2g_3 + m^0_7g_3a_2 - m^0_8g_2a_3 \\
 &+ m^0_9a_2a_3 - m^0_{10}a_5;\\
c_1={}&\sum_{i=1}^5 (m^1_ig_2+n^1_ia_2)F_i(\varphi)
 - m^1_6g_2^2  -m^1_7g_2a_2 +  m^1_8 a_2^2 + m^1_9a_4; \\
c_2={}&m^2_1g_3 +m^2_2a_3; \quad c_3=\sum_{i=1}^5m^3_iF_i(\varphi)
 - m^3_2g_2 +m^3_3a_2,
\end{split}
\end{equation*}
where
\[
F_i(\varphi) \equiv \left(\delta_{i,1} \beta_1+\delta_{i,2}h^2+
\delta_{i,3}\Big(\frac{\beta_1}{h'}\Big)^4+
\delta_{i,4}h \Big(\frac{\beta_1}{h'}\Big)^2+
\delta_{i,5}\Big(\frac{\beta_1h^{(4)}}{h'{}^2}\Big)  \right) .
\]
Here $m^j_i,\,n^j_i$ denotes some  numerical rational fractions,
$\beta_1=12\wp^2(\varphi)-g_2,\,h=\wp(\varphi)$,
$a_i=a_i(\wp(\varphi))$ denote some complicate functions,
the explicit form of
which we do not present here.

\subsection{Elliptic 3-gap solutions} \label{secCC}

The possible types of the time dependent elliptic 3-gap solutions of
the KdV equation (\ref{pot}) are determined by the values $N$ which are
obtained from the compatibility condition between (\ref{pot}) and
(\ref{U3}) as $t \to 0$. Under this  condition, the number $N$ takes values
$\overline {6,8}$ in the formula (\ref{pot}).

1. The values $N=6$ and $N=7$ determine two elliptic 3-gap solutions
of the KdV equations with initial conditions 1 and 2 of the system
(\ref{U3}), which have the form
\begin{equation}
U(z,t)=2\sum_{i=1}^6\wp(z-\varphi_i(t)),\quad{\mathrm {and}}\quad
U(z,t)=2\sum_{i=1}^7\wp(z-\varphi_i(t))-2\wp(\varphi_{07}),\label{U367t}
\end{equation}
respectively.
The time evolution of the poles $\varphi_i(t)$ are described by relations
(\ref{pd}) with the lower integration limits
$\varphi_{0,i}|_{i =\overline {1,6}}=0$ at the initial condition 1 and
$\varphi_{0,i}|_{i =\overline {1,6}}=0,\,\varphi_{0,7}=\omega_i$ at
the initial condition 2.

By  substituting the expressions (\ref{U367t}) into the finite-band equation
(\ref{b4}) and equating coefficients of the Laurent expansion
in $\wp$ of the right-hand sides to zero, lead to the two equations
\begin{equation}
\sum_{i=0}^6c_{6,i}X^i=0,\quad{\mathrm {and}}\quad\sum_{i=0}^7c_{7,i}X^i=0,
\label{X67}
\end{equation}
corresponding to the two 3-gap solutions with the values $N=6$ and
$N=7$, respectively. Coefficient functions $c_{6,i}$ and $c_{7,i}$ in
(\ref{X67}) are rational functions on $\wp(\varphi_i)$ and
$\wp'(\varphi_i)$. Therefore $X_i$ as roots of (\ref{X67}) are functions
of $\varphi_i$, i.e. $X_i=X_i(\varphi_i)$ where $i=\overline {1,6}$ and
$i=\overline {1,7}$ at initial conditions 1 and 2, respectively.

2. The value $N=8$  determines two similar elliptic 3-gap solutions of
the KdV equations with the initial conditions 3 and 4 of the system
(\ref{U3}), which have the form
\begin{equation}
U(z,t)=
2\sum_{i=1}^8\wp(z-\varphi_i(t))-2(\wp(\varphi_{07})+\wp(\varphi_{08})).
\label{U38t}
\end{equation}
The poles $\varphi_i(t)$ of (\ref{U38t}) are described by relations
(\ref{pd}), with the lower integration limits
$\varphi_{0i}|_{i =\overline {1,6}}=0,\quad
\varphi_{07}=\omega_i,\quad \varphi_{08}=\omega_j$
and $\varphi_{0i}|_{i =\overline {1,6}}=0,\quad
\varphi_{07}=-\varphi_{08}=-\varphi_3$ at initial conditions 3 an 4,
respectively.

Substituting the expressions (\ref{U38t}) into the finite-gap equation
(\ref{b4}) and equating the coefficients of the Laurent
expansion in $\wp$ of the right-hand sides to zero, we obtain
two equations of the form
\begin{equation}
\sum_{i=0}^8c_{6,i}X^i=0,
\label{X8}
\end{equation}
where $c_i=c_i(\varphi)$. Eight solutions of (\ref{X8}) coincide with
eight functions $X_i(\varphi_i),\,i=\overline {1,8}$.

The proposed approach is applicable for computing
arbitrary elliptic $n$-gap solutions of the KdV equation.
It can also be applied for
computing finite-gap elliptic solutions for other integrable
equations.

\section{Conclusion} \label{secD}

The solution of the KdV equation in the class of elliptic
finite-gap functions is reduced to the solution of the system of
finite-gap equations and auxiliary dynamic equations. In terms of
the elliptic $\wp$-function this system reduces to simple algebraic
relations which determin the parameters of the unknown solutions which are
represented as rational functions of the $\wp$-function. This approach gives
a simple algorithm for calculating arbitrary elliptic finite-gap
solutions of the KdV equations at an initial time, which was
demonstrated
by the example of 1-, 2- and 3-gap solutions. The time evolutions of the
unknown solutions with a known  $\wp$-functional representation, are determined
by the dynamics of their poles, which is described by coupled systems
of dynamic equations. The latter always can be integrated with the
help of the finite-gap equations. This was demonstrated by the example of
elliptic 1-, 2- and 3-gap solutions.

The above approach will also be applied to 
other
integrable nonlinear equations in a future paper.

\subsection*{Acknowledgements} This research was supported by the CDRF grant
UM1-325.

\label{korostil_lp}

\end{document}